\documentstyle[twocolumn,aps]{revtex}
\begin{document}

\title{Scaling Properties in Highly Anisotropic Systems}

\author{Qiming Li$^1$, S.~Katsoprinakis,$^2$, E.~N.~Economou$^2$,
C.~M.~Soukoulis$^{1,2}$}

\address{$^1$Ames Laboratory and Department of Physics and Astronomy,\\
Iowa State University, Ames, IA 50011\\}

\address{$^2$Research Center of Crete and Department of Physics, University
of\\
Crete, P. O. Box 1527, 71110 Heraklion, Crete, Greece\\}

\author{\parbox[t]{5.5in}{\small
Scaling of the conductances and the finite-size localization lengths is
generalized
to anisotropic systems and tested in two dimensional systems. Scaling
functions of isotropic systems are
recovered once the dimension of the system in each direction
is chosen to be proportional to the localization length. It is also
shown that the geometric mean of the localization lengths is a
function of the geometric mean of the conductivities. The ratio of the
localization lengths is proportional to the square root of the ratio
of the conductivities, which in turn is proportional to the anisotropy
strength t, in the weak scattering limit.
\\ \\
PACS numbers: 72.15.Rn \qquad 71.30.+h \qquad 74.25.Fy\\ }}

\maketitle

Scaling theories have been successfully applied to the problem of Anderson
localization\cite{lee,kramer}
involving the effects of disorder on the nature of the electronic wave
function.  The most remarkable result of the scaling theory is its
prediction\cite{scaling} that a continuous metal-insulator transition
exists in three dimensions, and all the states are localized in two dimensions,
in the absence of magnetic field and spin-orbit interaction.
The essential hypothesis of the one parameter scaling theory of localization
is that the rate of change of the conductance
when the size of the system changes is
controlled by the conductance alone. The critical conductance $G_c$
that separates
true metals from insulators is estimated\cite{lee}
to be $G_c = 0.1 \frac {e^2} {\hbar}$. The scaling function should also be
universal within a class that is characterized by a few general symmetries of
the governing Hamiltonian. The scaling theory results are supported
by a large number of numerical studies\cite{kramer} in d=2 and d=3. Most
notably,
finite size scaling calculations\cite{mack} on the transmission
properties of a quasi-one
dimensional system explicitly demonstrated the existence of a universal
scaling function close to the critical regime.

Most of the previous work involves isotropic systems. Recently, the
problem of Anderson localization in anisotropic systems has attracted
considerable attention\cite{wolfle,anis3d,xue,rice}, largely due to the fact
that a large variety of materials are highly anisotropic. It was recently
shown\cite{anis3d} that in a highly
anisotropic system of weakly coupled planes, states are localized in
the direction parallel and perpendicular to the plane at exactly
the same amount of critical disorder, in support of the one parameter
scaling theory which excludes the possibility of having a wave
function localized in one direction and extended in the other two.
However, several issues regarding the relation between the
conductances in different directions were raised. Most importantly,
the question of scaling of conductances and localization lengths
was not resolved\cite{anis3d}. Although anisotropy is known not 
to change the universality and thus the
critical behavior of the system\cite{privman}, the exact form of the
scaling function, on the other hand is expected to
depend on the anisotropy in
the form of anisotropic physical parameters such as anisotropic hopping
integrals or geometrical aspect ratios\cite{hu}.

Extending the scaling argument to an anisotropic system, we
assume that the logarithmic derivative, $\beta_i$, of
the dimensionless conductance $g_i$, in any direction will be a function of
the conductance in that direction as well as other directions,
\begin{equation}
\beta_i=\frac{d\log g_i}{d\log a}=\beta_i(\{g_i\})
\end{equation}
where $a$ is an appropriate length scale.
All the $g_i$ become relevant scaling parameters. All other physical
quantities, such as anisotropic hopping integrals or anisotropic geometrical
shapes, should enter only through the conductances $g_i$.
Exactly the same argument can be applied to the scaling function
of localization length, obtained from transfer matrix calculations with a
quasi-one dimensional geometry of cross section $M_j\times M_k$,
\begin{equation}
\frac {\lambda_i (M_j, M_k)}  {\xi_i} = h(\frac {M_j} {\xi_j},
\frac {M_k}{\xi_k})
\end{equation}
where $\lambda_i$ is the finite size localization length in the
direction $i$ and $\xi_l$ ($l$ = 1,2,3) is the localization length  
for the infinite system. The fundamental assumption
in Eq. (2) is that localization lengths provide the only characteristic 
length scale. Once the characteristic lengths are measured in
terms of the localization lengths in the corresponding directions, the
scaling behavior of the system within the same universality class,
are governed by the same equation.

The scaling functions $\beta_i$ and $h$ describe the behavior of
both systems with isotropic Hamiltonians but non-cubic geometry as
well as systems with anisotropic Hamiltonians. Scaling in anisotropic systems
 in general is not known. Only when the conductances in all directions are 
the same, then the scaling
function $\beta_i$ will be exactly the same as that of a cubic isotropic
system. For an anisotropic system, this can only be achieved by choosing an
appropriate geometry which might not be known a priori. As an example
we will see that indeed such a procedure
works in a system with highly anisotropic hopping. We will demonstrate that
in two dimensional systems, Eq.(2) can
be applied straightforwardly such that all the data are described by the
scaling functions of the isotropic system. Furthermore, we will
also show that the geometric mean of the localization lengths is a universal
function of the geometric mean of the bare conductivity, and their ratio
can also be estimated in the weak scattering limit. These results follow 
directly from applying the basic idea of scaling theory.

We consider the following Hamiltonian for an anisotropic 2d disordered model
\begin{equation}
H = \sum_n \epsilon_n |n> < n| + \sum_{n,m} t_{nm} | n > < m|
\end{equation}
where n labels the sites of a square lattice.  The on-site energies
$\epsilon_n$ are independently distributed at random, within an interval
of width W.  The second term is taken over all pairs of nearest
neighbor (n.n.) sites, and the hopping integral $t_{nm}$=1 or t ($<$1),
depending on hopping directions.  As a convention, we have assigned the
direction with the large ($t_{nm}=1$)
and the small ($t_{nm} =t$) hopping value
as the parallel ($\parallel$) and the perpendicular ($\perp$)
directions, respectively.

In two dimensional systems, Eq. (2) can be written as
\begin{equation}
\frac {\lambda_i(M_j)} {M_j}  = \frac {\xi_i} {\xi_j} f(\frac {M_j} {\xi_j})
\end{equation}
where f(x)=h(x)/x is the scaling function for isotropic
systems. We have used the transfer matrix method\cite{kramer} to calculate
the finite size localization length $\lambda_i (M_j)$ for many $M_j$
 ($i$,$j$ = 1,2) (M=24, 48, 96, 120, 150, 300) and W=2-14
and several t and E, for both directions.
Figure 1 shows that all of our raw numerical data for both
$\lambda^\parallel_M$ and $\lambda^\perp_M$ for different anisotropies t
different disorder W and different energies E follow one universal curve,
by appropriately choosing the localization
length in the two directions, $\xi_\parallel$ and $\xi_\perp$.
The solid line through the data in Fig. 1, is the 2d isotropic
scaling function.
This is a direct confirmation of the scaling relation Eq. (4).

An important consequence of Eq. (4) is that at the critical point,
if any, the geometric mean of the ratio
of the finite-size localization length
to the cross-section width is a constant. This was indeed found\cite{anis3d}
to be true
but interpreted instead as a result of possible conformal invariance.
We point out that at the critical point,the geometric mean of the
conductances along the different directions may not be a constant.
This behavior of the conductances is different from that of
$\lambda_{M}/M$ and needs further study for its complete understanding.

To further test the scaling idea we have calculated the
conductance G in the two different directions for our anisotropic system.
From the Multichannel Landauer formula\cite{pichard,g}, $G=\frac{e^2} {h}
Tr(t^\dagger t)$, where t is the transmission matrix. 
With anisotropic hoppings, one should choose a geometry other than
the square such that the conductance is the same in all the directions and
then scale up the size of the system\cite{rice}. The
conductance should remain isotropic
if one parameter scaling theory is correct\cite{wolfle}.  We have tested
this idea
in a 2d system with
t = 0.1. The ratio of the two localization lengths was found to be
10 at W =
3.6.  We have scaled up the system of a rectangle of size M$\times$ N by
a factor 4, and from Fig. 2,
one can clearly see that although the conductance becomes extremely small
it remains isotropic, in agreement with the predictions of the one parameter
scaling theory\cite{wolfle}. For a square geometry and with the same
parameters as in Fig. 2 , the conductances in the two directions would 
diverge rapidly as the system size scales up.

Another length rescaling aspect can be seen by considering the self-consistent
theory of localization. It was shown in an earlier work\cite{anis3d} that in
order
 the localization criteria to be the same in all directions,
the length scale has to be chosen proportional to the square root of
the bare conductivity. This leads to an equation for the metal-insulator
transition that is exactly the same as that of the isotropic system
except that both the bare conductivity and the effective lattice constant
are replaced by their geometric means. A direct consequence of this
formulation is that the geometric mean of the localization (or correlation)
lengths should be only a function of the geometric mean of the
bare conductivities, ie,
\begin{equation}
<\xi>_g = f_l (<\sigma_0>_g)/S_f,
\end{equation}
where $<>_g$ denote the geometric mean of the values in the
two directions.  $\sigma_0$ is the bare conductivity and
$S_f$ is the Fermi surface area that enters through
the relation $\sigma_0 \sim S_f \ell$. $\ell$ is the mean free path.
$f_l$ is a function that can be obtained via the Potential Well
Analogy(PWA) or the self consistent theory of localization.
Using the PWA Ref.13 obtained $\xi = 2.72 \ell
exp [\pi^2 \hbar \sigma_0 /e^2]$.

Eq. (5) can be easily checked in the weak disorder limit, at which the
geometric mean of the bare conductivity can be shown to be
$\bar \sigma_0 = 15 {\protect \sqrt 2 t /\pi W^2}$, within the
Coherent Potential
Approximation (CPA)[14].  For the 2d anisotropic system,
$S_f(E=0) = 4\pi {\protect{\sqrt {1 + t^2}}}$. We have plotted
$S_f<\xi_\perp\xi_\parallel>^{1/2}$ versus $15 {\protect \sqrt 2 t /\pi W^2}$
, and find that the data fall into one universal curve for
all the different anisotropies t and disorder W.
This weak scattering limit behavior of the
geometric mean of the localization lengths versus the geometric mean of the
conductivities is very suggestive of the way the localization lengths have
to scale.  The full expression, valid for all disorder strength,
\begin{equation}
\sigma_{i0} = {2e^2\hbar \over \pi} \sum_k v^2_i(k) {\Sigma^2_2\over
{[ (E -
\Sigma_1 - E(k))^2 + \Sigma^2_2]^{2}}}
\end{equation}
can also be evaluated. $\Sigma = \Sigma_1 - i\Sigma_2$ is
the self-energy obtained by solving a self-consistent
equation\cite{wolfle,loc,green}.
This shows remarkable good scaling as shown in Fig. 3, including results
for E=0, as well as for $E\neq 0$.
The  curve in Fig. 3 shows how the geometric mean of the
localization length depends on the geometric mean of the bare conductivity
in a universal fashion, independent of the anisotropy, energy and disorder.
These results are a strong confirmation of scaling in anisotropic systems.
Notice that the geometric mean of the conductivities $<\sigma_0>_g$,
and not of the conductances, is the appropriate quantity that gives the
same results as in the isotropic case. It is therefore appropriate
that $<\sigma_0>_g$
will be used in the interpretation of experiments in highly anisotropic
systems.

The ratio of the localization lengths can be obtained by carrying the length
rescaling idea further.  We can see that the conductances in all the directions
should be the same, if the dimension of the system is proportional to the
localization length in that direction. This implies the following relation
\begin{equation}
\frac {\xi_i} {\xi_j} = (\frac {\sigma_i} {\sigma_j})^{1/2} =
(\frac {\sigma_{i0}} {\sigma_{j0}})^{1/2}  (\frac {\alpha_i} {\alpha_j})^{1/2}
\end{equation}
$\sigma_i$ is the exact value of the conductivity, $\sigma_{i0}$ is the bare
conductivity which can be calculated within the CPA, $\alpha_i$ is the
correction
of the bare conductivity in the i direction. It is very difficult to calculate
the correction $\alpha_i$, but it approaches one in the weak scattering limit.
In Fig. 4, we show the  results of
$\xi_\perp / t\xi_\parallel$ versus 1/W for different anisotropies t and
energies E. In the weak disorder limit, we can approximate $\sigma$ by
$\sigma_0$,
and this is shown as open symbols in Fig. 4.
Notice that in the weak disorder limit, W $\rightarrow$ 0,and for t
$\rightarrow$ 0, $\sigma_{0\perp} / \sigma_{0\parallel} \sim t^{2}$ and by
using Eq. (7),one obtains for the ratio of the localization lengths
, $\xi_\perp / \xi_\parallel \sim t$. This behaviour is clearly seen in
Fig. 4 for large 1/W.
Agreement with the CPA results for the conductivity are
excellent for weak disorder. Deviation of the ratio from the open symbols
for strong disorder indicates that the true conductivity at length
scale $\xi$ is stongly normalized compared with the bare conductivity.
However, it is notable that the trend of the ratio as W increases is captured
by the simple expression. For large W, no dependence on E should
be expected for small E, thus the ratios converge to the same value for
different E with t=0.3, as can be seen in the insert in Fig. 4.

In summary, we have performed an extensive numerical study of the scaling
properties
of highly anisotropic systems. Scaling functions of isotropic systems
are recovered once the dimension of the system in each direction is chosen
to be
proportional to the localization length.In the localized regime, the ratio
of the localization lengths is proportional to the square root of the ratio
of the conductivities which in turn is proportional to the strength of the
anisotropy t (i.e. $\xi_\perp / \xi_\parallel \sim t$ ).
Recall that in the extended regime\cite{wolfle,anis3d} the ratio of the
correlation lengths is proportional to the ratio of the conductivities
(i.e. $\xi_\perp / \xi_\parallel$ = $\sigma_{0\perp} / \sigma_{0\parallel}
\sim 1/t^{2}$ ). It was also shown that the
geometric mean of the localization lengths
is a function of the geometric mean of the
conductivities, not of the conductances.
Finally, it was numerically shown that the
conductances along the two different
directions of the anisotropic system are the same, provided
that the dimension of the anisotropic system is proportional to
the localization length in this direction.
This procedure can be easily used in other anisotropic systems.

Ames Laboratory is operated for the U.S. Department of Energy by Iowa
State University under Contract No. W-7405-Eng-82.
This work was supported by the Scalable Computing Laboratory
of Ames Lab, the director for Energy
Research, Office of Basic Energy Sciences, NATO Grant No. CRG 940647. This
work was also supported in part by  EU grants and a $\Pi$ENE$\Delta$ Research Grant of
the Greek Secretariat of Science and Technology.

\begin{figure}
\caption{The numerically determined scaling function for the
2D anisotropic system for different anisotropic constants t, different
energies E and disorder W. The solid line through the data is the 2D isotropic
scaling function. The y-axis is $\xi_j\lambda_i(M_j)/\xi_iM_j$,
while the x-axis is $M_j/\xi_j$. The index i and j can be either the parallel
or the perpendicular direction, respectively.
\label{1}}\end{figure}

\begin{figure}
\caption{The conductance G in units of $e^2/h$ of an anisotropic system M x N,
versus M for t=0.1 and E=0. Notice that G along the two directions is exactly
the same.
\label{2}}\end{figure}

\begin{figure}
\caption{The product of the Fermi surface $S_f$ with the geometric mean of
the localization lengths $<\xi_\perp \xi_\parallel>^{1/2}$ is plotted versus
the geometric
mean of the bare
conductivities $<\sigma_{0\perp} \sigma_{0\parallel}>^{1/2}$ for all the
energies E, t's and W.
\label{3}}\end{figure}

\begin{figure}
\caption{The ratio of $\xi_\perp/t\xi_\parallel$ is plotted versus 1/W, 
for t=0.1, 0.3 and 0.6 with E=0. $\xi_\perp$ and $\xi_\parallel$
are the localization lengths along the two propagating directions.
 The solid symbols are the numerical results, while the open symbols
are the CPA results.
In the insert the numerical results of $\xi_\perp/ t\xi_\parallel$ versus 1/W is plotted for t=0.3 with
E=0.0, 1.5 and 2.0.
\label{4}}\end{figure}


\begin{references}
\bibitem{weak} In the weak scattering limit, the static conductivity
$\sigma_{i0}$ is given
[6,13] by $\sigma_{i0} = e^2 S < v^2_i/v > \tau/4\pi^2$, where i = x and y
and S(E) is
the area of the surface $E(\vec k) = E = 2 cos k_x a + 2t cos k_y \alpha$
and the average
is taken over this surface.  Normally S(E) is tedious to calculate, however
for E = 0,
one obtains $S(E=0) = 4 \pi {\protect \sqrt{1+t^2}}$.  The S(E) $< v^2_i/v>$
can be
expressed [6,13] in terms of the lattice Green's function.  For the
anisotropic system we
have $S<v_x^2/v > = 8 \pi [ImG(E;0,0) - ImG(E;2,0)]$ and
$S<v^2_y/v> = 8 \pi t^2 [ImG(E;0,0) - ImG(E,0,2)]$.  For E = 0
we can show [13] that ImG (0;0,0) =
K(t)/$\pi$,
ImG(0;0,2) = (2-t$^2$)K(t)/$\pi t^2$ - 2E(t)/$\pi t^2$ and ImG (0;2,0) =
K(t)/$\pi$ -
2E(t)/$\pi$, where K and E are the complete elliptic integrals of the
first and second kind respectively.  In the limit of small coupling t
$\rightarrow$ 0, we obtain that $S < v^2_x/v > \simeq 8 \pi/\hbar$ and $S<
v^2_y/v >
\simeq 4\pi t^2 /\hbar$.  By using these limiting values, we obtain the
geometric value
$\bar \sigma_0 = (\sigma_{0x} \sigma_{0y})^{1/2}$ of the two conductivities
is equal to
$\bar \sigma_0 = 15 {\protect \sqrt 2 t /\pi W^2}$, where we have used
the fact that
the relaxation time $\tau = \hbar /2 \Sigma_2 =15 \hbar /W^2$ in the weak
scattering limit.
\end{references}
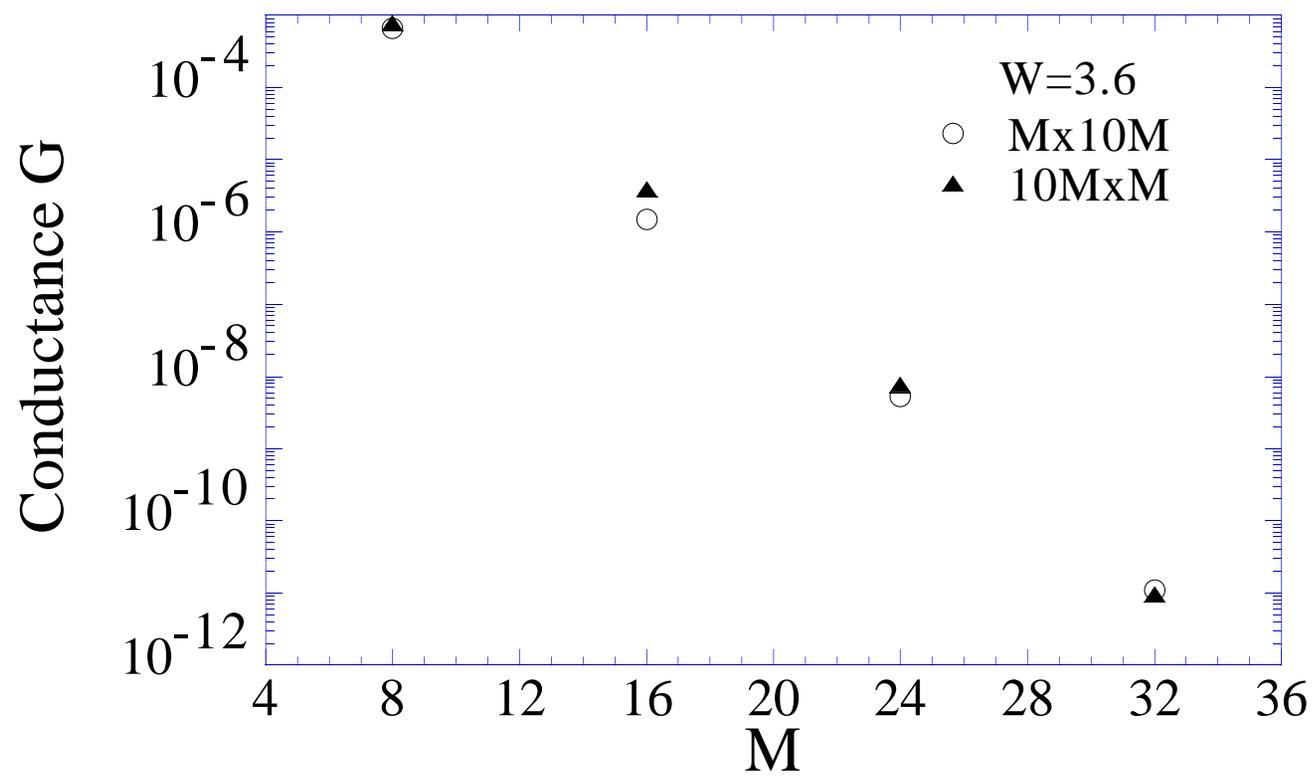
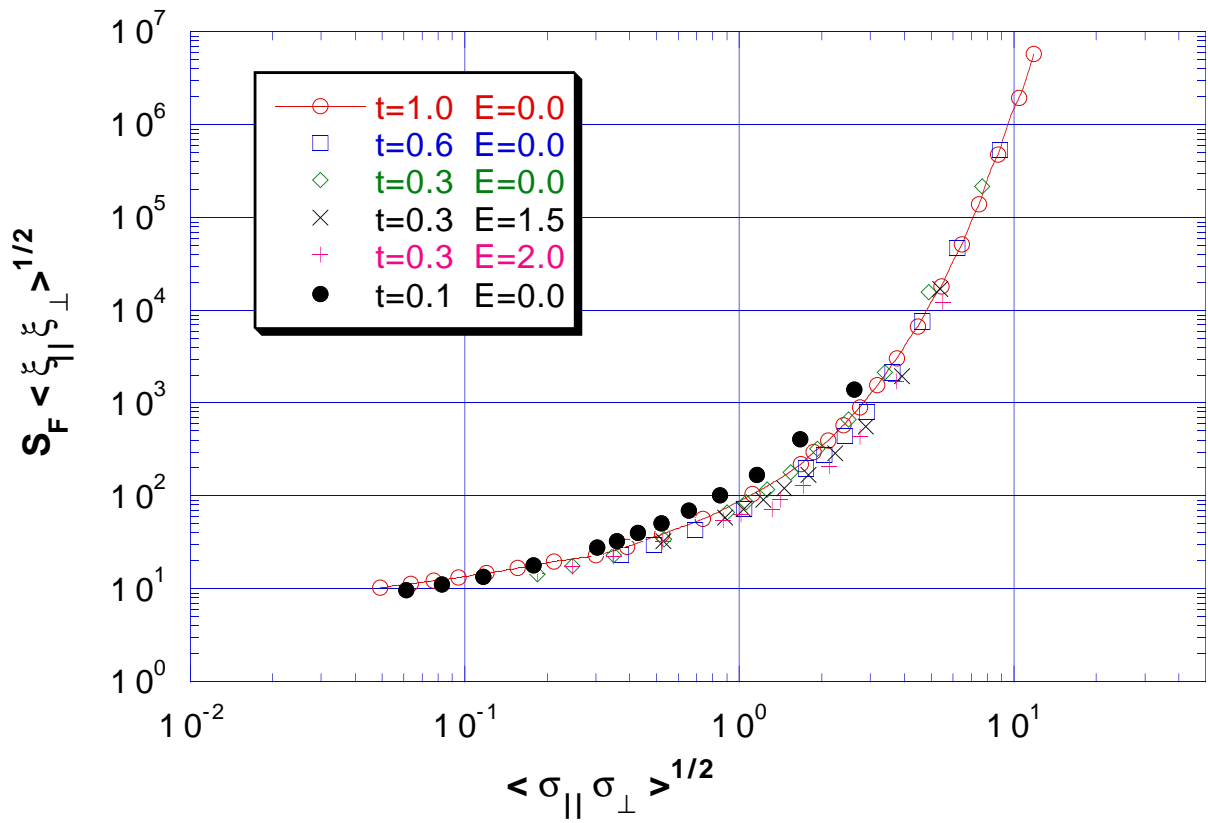
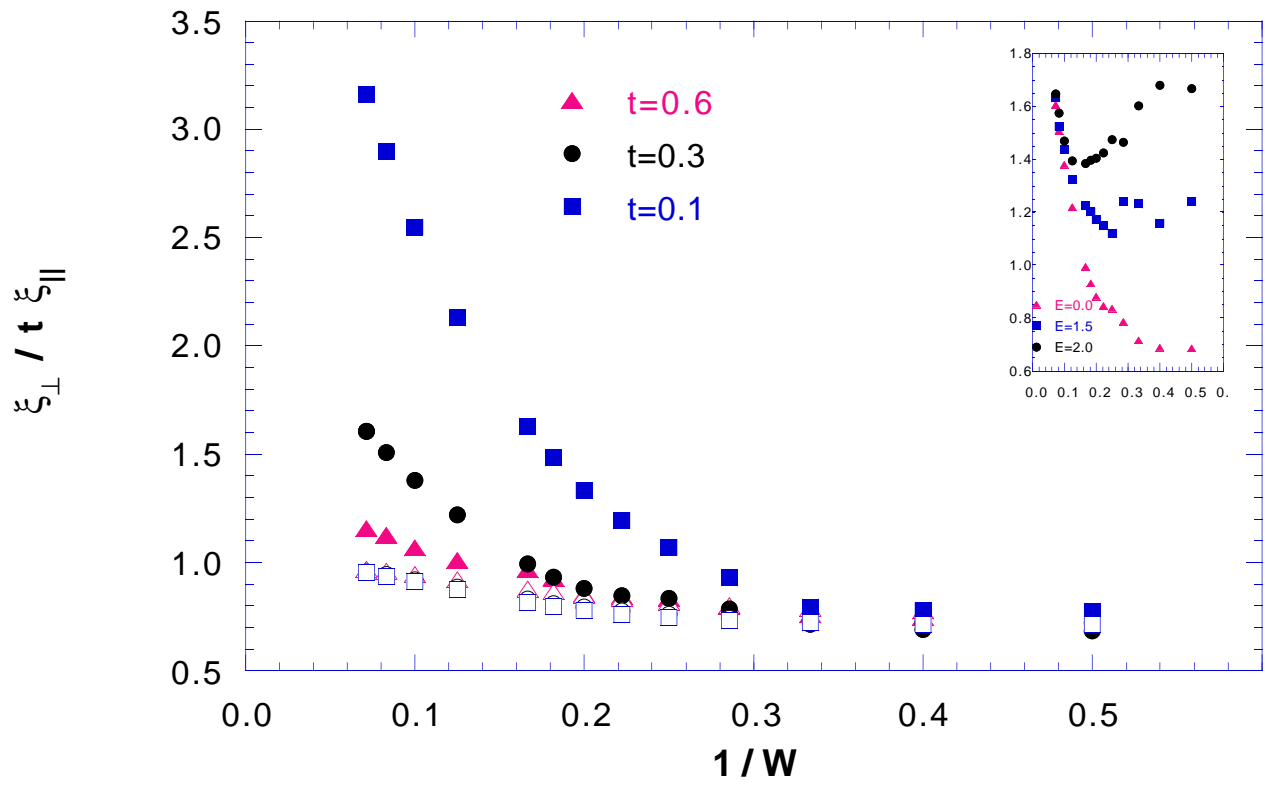
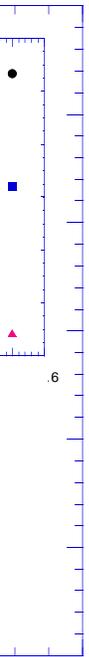
\end{document}